\documentclass[english,aps,preprint,superscriptaddress]{revtex4}
\usepackage[]{fontenc}
\usepackage[latin9]{inputenc}
\usepackage{amsmath}
\usepackage{esint}  %%???
\usepackage{epsfig}
\usepackage{graphicx}
\usepackage{bm}% bold math

\makeatletter
\@ifundefined{textcolor}{}
{%
 \definecolor{BLACK}{gray}{0}
 \definecolor{WHITE}{gray}{1}
 \definecolor{RED}{rgb}{1,0,0}
 \definecolor{GREEN}{rgb}{0,1,0}
 \definecolor{BLUE}{rgb}{0,0,1}
 \definecolor{CYAN}{cmyk}{1,0,0,0}
 \definecolor{MAGENTA}{cmyk}{0,1,0,0}
 \definecolor{YELLOW}{cmyk}{0,0,1,0}
 }

\makeatletter

\newcommand{\be}{\begin{equation}}
\newcommand{\ee}{\end{equation}}
\newcommand{\ba}{\begin{align}}
\newcommand{\ea}{\end{align}}
\def\bea{\begin{eqnarray}}
\def\eea{\end{eqnarray}}

\usepackage{slashed}

\makeatother

\usepackage{babel}

\begin{document}

\title{Bound on Noncommutative Standard Model with Hybrid Gauge Transformation via Lepton Flavor Conserving $Z$ Decay }

\author{Weijian Wang}
\affiliation{Zhejiang Institute of Modern Physics, Department of Physics, Zhejiang
University, Hangzhou 310027, P.R. China}
\affiliation{Department of Physics, North China Electric Power
University, Baoding, P. R. China}

\author{Jia-Hui Huang}
\affiliation{Center of Mathematical Science, Zhejiang University,
Hangzhou 310027, P.R.China}

\author{Zheng-Mao Sheng}\thanks{corresponding author's Email: zmsheng@zju.edu.cn}
\affiliation{Institute for Fusion Theory and Simulation, Department of Physics, Zhejiang University, Hangzhou 310027,
P.R. China}
\affiliation{Zhejiang Institute of Modern Physics, Department of Physics, Zhejiang
University, Hangzhou 310027, P.R. China}

\begin{abstract}

The $Z\rightarrow e^{+}e^{-}$ decay is studied basing on
the noncommutative standard model (NCSM) with the hybrid gauge
transformation. It is shown that if the latter is not included, the noncommutative correction to the amplitude of the $Z\rightarrow e^{+}e^{-}$ appears only as a phase factor, so that there is no new physical effect on the decay width.
However, when the hybrid gauge transformation is included, the noncommutative effect appears in the two-body decay process. The discrepancy between the experimental branch ratio and the standard model prediction allows us to set the bound on the noncommutative parameters.

\end{abstract}
\maketitle

The concept of noncommutative (NC) space-time was firstly introduced
by Snyder in 1947\cite{snyder1947}. Interest on NC space-time was revived since
it appeared in the string theory and other quantum gravity models as effective theories in low energy limit\cite{seiberg1999, Connes1998, Douglas2001, Szabo2003}. In a popular NC
model the NC space-time is characterized by a coordinate operator satisfying
\begin{equation}
[\hat{x}_{\mu},\hat{x}_{\nu}]=i\theta_{\mu\nu}=\frac{ic_{\mu\nu}}{\Lambda^2_{NC}},
\label{homer}
\end{equation}
where $\theta_{\mu\nu}$ is a constant antisymmetric matrix. Its elements have a
dimension of $(mass)^{-2}$. Here $c_{\mu\nu}$ is a real antisymmetric
matrix, whose dimensionless elements are assumed to be
of order unity, and the NC scale $\Lambda_{NC}$ characterizes the threshold where the NC effect
becomes relevant and its role can be compared to that of ${\hbar}$ in
quantum mechanics. The existence of a finite $\Lambda_{NC}$ implies the existence
of a fundamental space-time distance below which
the space-time coordinates become fuzzy. By Weyl-Moyal correspondence, the
quantum field theory in NC space-time is equivalent to that in
ordinary space-time with the normal product of the field variables
replaced by the star product, defined by \cite{moyal}
\begin{equation}
\phi_{1}(x)*\phi_{2}(x)=\exp\left(\frac{i}{2}\theta^{\mu\nu}\partial_{\mu}^{x}\partial_{\nu}^{y}\right)\phi_{1}(x)\phi_{2}(y)|_{y\rightarrow
x}.\label{eq:moyalproduct}\end{equation}
Using this method, a noncommutative extension of the standard
model (NCSM) has been proposed\cite{Calmet1}, where the SU(N) Lie
algebra is generalized to the enveloping algebra via the
Seiberg-Witten map (SWM) \cite{seiberg1999}. The SWM
is a map between the noncommutative field and its counterpart
in ordinary space-time as a power series of the NC parameter
$\theta_{\mu\nu}$. The NCSM predicts the NC-corrected particle vertex
and many new interactions beyond the standard model, for instance,
the $Z-\gamma-\gamma$ and $Z-g-g$ vertices. The rich phenomenological
implication has been intensively examined in high energy
processes for possible experimental signal or give a
bound on the noncommutative scale $\Lambda_{NC}$ \cite{Das08,Wang2012}.

On the other hand, the neutrino oscillation experiments have shown
convincing evidence of massive neutrinos and leptonic favor mixing
\cite{neu2002,exact2011}, so that in constructing NCSM the neutrino
mass should be included. It is found that the hybrid gauge
transformation and hybrid SWMs are needed to accommodate the seesaw
mechanism \cite{exact2011}. The most popular mechanism
for generating neutrino mass, and the gauge invariance of NC gauge
theory. The hybrid gauge transformation and hybrid SWM have been
adopted in the Higgs sector of NCSM to ensure covariant Yukawa terms
\cite{Calmet1}. In this scenario, the Higgs fields feel a ``left"
charge and a ``right" charge in the NC gauge theory and transforms
from left side and right side correspondingly.
Although it is only applied to the Higgs sector in Ref.\
\cite{Calmet1}, in Ref.\ \cite{exact2011} it was shown that this
method can in principle be extended to consider fermion fields. A new
physics predicted by the hybrid gauge transformation is the
tree-level coupling between the neutrino and the photon
\begin{equation}
i\kappa e(\hat{A}_{\mu}*\nu-\nu*\hat{A}_{\mu}),
\label{arb}\end{equation}
where $\hat{A}$ and $\hat{\nu}$ are the photon and neutrino fields,
respectively. In NCQED, to maintain the gauge invariance the charge
is quantized to  -1, 0, and 1, corresponding to the interaction
terms $e\hat{A}_{\mu}*\psi$, $e(\hat{A}_{\mu}*\psi-\psi*\hat{A}_{\mu})$, and $e\psi*\hat{A}_{\mu}$.
However, in the NCSM based on the enveloping algebra, the Seiberg-Witten
map can overcome the constraint of charge quantization and guarantee
the gauge invariance at the same time. So one can loosen the
constraint on charge quantization and arbitrarily set the electric charge in
\eqref{arb} as $\kappa e$. The photon-neutrino
interaction can lead to interesting phenomena, and has been
discussed by many authors (see Ref.\ \cite{Trampetic2012} and the
references therein).

It is interesting to see if the hybrid gauge transformation will lead to
other phenomenological effects. In Ref.\ \cite{zvv}, the anomalous
$Z-\nu-\nu$ interaction is derived and the invisible $Z$ decay process
$Z\rightarrow \nu\bar{\nu}$ is studied. It is shown that for
$\kappa=1$, the current experimental result
$\Gamma_{\textup{invisible}}=(499.0\pm1.5)$ MeV \cite{PDG} allows us
to set the bound $\Lambda_{NC}\geq 140$ GeV on the noncommutative scale.

Besides the invisible decay, it is also of interest to investigate
the $Z\rightarrow l^{+}l^{-}$ channel. In the standard model, the $Z$
boson decays into lepton pairs through the lepton flavor conserving
(LFC) interaction at the tree level. Up to now, the current
experimental data produces $\textup{Br}(Z\rightarrow
e^{+}e^{-})=3.363\pm0.004\%$, $\textup{Br}(Z\rightarrow
\mu^{+}\mu^{-})=3.366\pm0.007\%$ and $\textup{Br}(Z\rightarrow
\tau^{+}\tau^{-})=3.370\pm0.0023\%$\cite{PDG}. On the other hand,
the theoretical prediction from SM, including the loop correction, is
$\textup{Br}(Z\rightarrow e^{+}e^{-})=\textup{Br}(Z\rightarrow
\mu^{+}\mu^{-})=3.3346\%$ and $\textup{Br}(Z\rightarrow
\tau^{+}\tau^{-})=3.3338\%$\cite{Chong}. In this paper, we focus on
the $Z\rightarrow e^{+}e^{-}$ and in the following
calculation the zero lepton mass approximation is adopted. The gap
between the experimental results and the theoretical prediction is of
order $0.03\%$ and exhibits possible existence of new physics
beyond the standard model. Motivated by this, various models beyond
the SM have been discussed\cite{Chong,De,ILT}. In
Ref.\ \cite{ILT}, the same issue has been discussed in the NCSM framework without
the hybrid gauge transformation. However, our detailed analysis\cite{Wang2012} showed that the NC effect only appears in the $Z-l-l$ vertex as a phase factor, so that no
physical deviation appears. Here we study $Z\rightarrow
e^{+}e^{-}$ in the frame work of NCSM with hybrid gauge
transformations. From the viewpoint of gauge invariance, the hybrid
feature also effects the charged lepton interaction. To see
this, we briefly review our earlier
results\cite{Wang2012}. The action of lepton in NCSM can be written as
\begin{equation}
\hat{S}_{lepton}=i\int d^4x[\bar{\hat{\Psi}}_{L}\gamma^{\mu}D_{\mu
L}\hat{\Psi}_{L}+\bar{\hat{l}}_{R}\gamma^{\mu}D_{\mu R}\hat{l}_{R}]
\label{lepton}\end{equation}
with $\hat{\Psi}_{L}$ and $\hat{l}_{R}$ denoting the doublet lepton
\begin{equation}
\hat{\Psi}_{L}=\begin{pmatrix}{\hat{\nu}_{L}}\\{\hat{l}_{L}}\end{pmatrix},
\end{equation}
and the right-handed singlet lepton, respectively. Under the hybrid gauge
transformation, the $\hat{\Psi}_{L}$ and $\hat{l}_{R}$ transform as
\begin{equation}\displaystyle\begin{split}
&\delta_{\hat{\Lambda}}\begin{pmatrix}{\hat{\nu}_{L}}\\{\hat{l}_{L}}\end{pmatrix}
=ig_{Y}\left[(-\frac{1}{2}+\kappa)\hat{\Lambda}*\begin{pmatrix}{\hat{\nu}_{L}}\\{\hat{l}_{L}}\end{pmatrix}
-\kappa\begin{pmatrix}{\hat{\nu}_{L}}\\{\hat{l}_{L}}\end{pmatrix}*\hat{\Lambda}\right],\\
&\delta_{\hat{\Lambda}}\hat{l}_{R}=ig_{Y}[(-1+\kappa)\hat{\Lambda}*\hat{l}_{R}-\kappa\hat{l}_{R}*\hat{\Lambda}],
\end{split}\end{equation}
where $\hat{\Lambda}$ is the gauge parameter. Under the gauge
transformation above, the covariant derivatives in
Eq.\ \eqref{lepton} is
\begin{equation}
D_{\mu
L}\hat{\Psi}_{L}=\partial_{\mu}\hat{\Psi}_{L}-ig_{L}\hat{A}_{\mu}^{a}T^{a}*\hat{\Psi}_{L}
-(-\frac{1}{2}+\kappa)g_{Y}\hat{B}_{\mu}*\hat{\Psi}_{L}+i\kappa
g_{Y}\hat{\Psi}_{l}*\hat{B}_{\mu}, \label{c1}\end{equation}
\begin{equation}
D_{\mu R}\hat{l}_{R}=\partial_{\mu}\hat{l}_{R}-i\kappa
g_{Y}B_{\mu}*\hat{l}_{R}+i\kappa g_{Y}\hat{l}_{R}*B_{\mu},
\label{c2}\end{equation}
where $\hat{A}_{\mu}^{a}$ and $g_{L}$
are the $SU(2)_{L}$ gauge fields and a coupling constant. To get the
appropriate particle vertex, we should replace the fermion and gauge
fields in Eqs.\ \eqref{lepton}, \eqref{c1}, and \eqref{c2} by their
classical counterparts via appropriate Seiberg-Witten maps. The
detailed formation of Seiberg-Witten map is given in
Ref.\ \cite{exact2011}, where the so-called $\theta-$ exact formation
is adopted to include the contribution of all $\theta$ orders.
From the deformed Lagrangian\cite{Wang2012} one can then
obtain the Feynman rule of the $Z-l-l$ interaction
\begin{equation}
\frac{ie}{\sin2\theta_{W}}\gamma^{\mu}(C_{V}-C_{A}\gamma^{5})e^{\frac{i}{2}p_{1}\theta
p_{2}}+\frac{2\kappa
e\sin\theta_{W}}{\cos\theta_{W}}\gamma^{\mu}\sin(\frac{1}{2}p_{1}\theta
p_{2}), \label{rule}\end{equation}
where $p_{1}$ ($p_{2}$) is the
ingoing (outgoing) lepton momentum,
$p_{1}\theta p_{2}\equiv p_{1}^{\mu}\theta_{\mu\nu}p_{2}^{\nu}$,
$C_{V}=-\frac{1}{2}+2\sin^2\theta_{W}$, $C_{A}=-\frac{1}{2}$, and
$\theta_{W}$ denotes the Weinberg angle. We have applied the
equation of motion to the electron external line and omitted the vanishing terms
due to the on-shell condition.

Using the Feynman rule in Eq.\ \eqref{rule}, the derivative decay
width of $Z\rightarrow e^{+}e^{-}$ can be easily obtained in the $Z$
boson rest frame
\begin{equation}
\frac{d\Gamma}{d\cos{\theta} d\phi}=\frac{M_Z}{48\pi^2
}\Big[\frac{e^2}{\sin^2{2\theta_W}}(C^2_V+C^2_A)+(4\kappa^2
e^2\tan^2{\theta_W}-2\kappa
C_V\frac{e^2}{\cos^2\theta_W})\sin^2({\frac{1}{2}p_1\theta
p_2})\Big].\label{wid}
\end{equation}

In the calculation, we omit the lepton mass. As mentioned, the NC
parameter $\theta_{\mu\nu}$ is a fundamental constant that breaks the
Lorentz symmetry. Following the method adopted in Ref.\
\cite{Fu2007}, one can decompose $\theta_{\mu\nu}$ into two types:
the electric-like components
$\theta_{E}=(\theta_{01},\theta_{02},\theta_{03})$ and the
magnetic-like components
$\theta_{B}=(\theta_{23},\theta_{31},\theta_{12})$. Both of them are
assumed to be directionally fixed in a primary, unrotated reference.
That is, when discussing phenomena in the laboratory frame, the
Earth's rotation should be included. Defining $(\hat{X}, \hat{Y},
\hat{Z})$ to be the orthonormal basis of this primary frame,
$\theta_{E}$ and $\theta_{B}$ are
\begin{equation}
\theta_{E}=\frac{1}{{\Lambda_{E}^{2}}}(sin{\eta_{E}}cos{\xi_{E}}\hat{X}+sin{\eta_E}sin{\xi_{E}}\hat{Y}+cos{\eta_E}\hat{Z}),
\label{theE}\end{equation}
\begin{equation}
\theta_{B}=\frac{1}{{\Lambda_{B}^{2}}}(sin{\eta_{B}}cos{\xi_{B}}\hat{X}+sin{\eta_B}sin{\xi_{B}}\hat{Y}+cos{\eta_B}\hat{Z}),
\label{theB}\end{equation} where $\eta$ and $\xi$ denote the polar
angular and azimuth angular of NC parameter with $0\leq\eta\leq\pi$
and $0\leq\xi\leq2\pi$, respectively. Since we are in the
$(\hat{x},\hat{y},\hat{z})$ frame on Earth, it is necessary to find
an appropriate transformation matrix correlating the primary and
laboratory reference frames. Following Ref.\ \cite{Kam2007}, we have
\begin{eqnarray}
\left(\!\!\!\begin{array}{ccc}
{\hat{X}}\\{\hat{Y}}\\{\hat{Z}}\end{array}\!\!\!\right)=\left(\!\!\!\begin{array}{ccc}
c_{a}s_{\zeta}+s_{\delta}s_{a}c_{\zeta}&c_{\delta}c_{\zeta}&s_{a}s_{\zeta}-s_{\delta}c_{a}c_{\zeta}\\
-c_{a}s_{\zeta}+c_{\delta}s_{a}s_{\zeta}&_{\delta}s_{\zeta}&-s_{a}c_{\zeta}-s_{\delta}c_{a}s_{\zeta}\\
-c_{\delta}s_{a}&s_{\delta}&c_{\delta}c_{a}\end{array}\!\!\!\right)\left(\!\!\!\begin{array}{ccc}
{\hat{x}}\\{\hat{y}}\\{\hat{z}}\end{array}\!\!\!\right),
\label{tran}\end{eqnarray}
where the abbreviation $c_{\alpha} = \cos{\alpha}$ and $s_{\alpha} = \sin{\alpha}$, with $\alpha = a$, $\delta$ and $\zeta$ respectively, are used. Here, $\delta$ and $a$ define the location and
orientation of the experiment site, with
$-\frac{\pi}{2}\leq\delta\leq\frac{\pi}{2}$ and $0\leq a \leq 2 \pi
$, $\zeta=\omega t$ is the rotation angle, and
$\omega=2\pi/{23h56m4.09s}$ is the earth's angular velocity. Ignoring the Earth's revolution, the
collider machine returns to its original position after one day. Using Eqs.\ \eqref{theE}, \eqref{theB}, and \eqref{tran}, we get
\begin{equation}\begin{split}
&p_{2}\theta p_{1}=-\frac{s}{2\Lambda^{2}_{NC}}
(\sin\theta\cos\phi\Theta_{E}^{x}+\sin\theta\sin\phi\Theta_{E}^{y}+\cos\theta\Theta_{E}^{z})
\end{split}\label{zzz}\end{equation}
with
\begin{equation}\begin{split}
&\Theta_{E}^{x}=s_{\eta}c_{\xi}(c_{a}s_{\zeta}+s_{\delta}s_{a}c_{\zeta})+
s_{\eta}s_{\xi}(-c_{a}s_{\zeta}+c_{\delta}s_{a}s_{\zeta})-c_{\eta}c_{\delta}c_{a},\\
&\Theta_{E}^{y}=s_{\eta}c_{\xi}c_{\delta}c_{\zeta}+s_{\eta}s_{\xi}c_{\delta}c_{\zeta}+c_{\eta}c_{\delta},\\
&\Theta_{E}^{z}=s_{\eta}c_{\xi}(s_{a}s_{\zeta}-s_{\delta}c_{a}c_{\zeta})+s_{\eta}s_{\xi}(-s_{a}c_{\zeta}-s_{\delta}c_{a}s_{\zeta})
+c_{\eta}c_{\delta}c_{a}.
\end{split}\end{equation}

Substituting Eq.\ \eqref{zzz} into Eq.\ \eqref{wid}, we obtain the decay
width of $Z\rightarrow e^{+}e^{-}$ in the laboratory frame.

\begin{figure}
 \includegraphics[scale = 0.56]{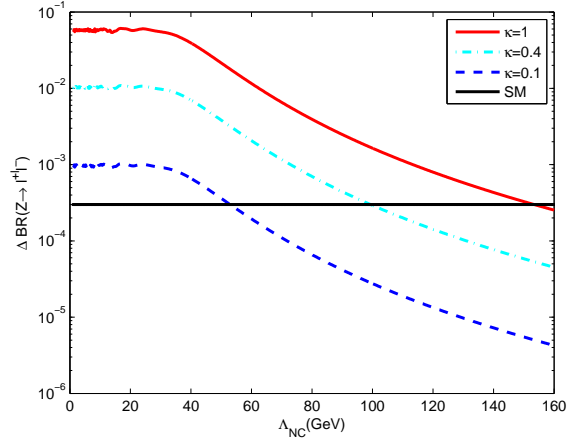}%
\caption {NC correction of the branch ratio $Z\rightarrow e^{+}e^{-}$
as a function of $\Lambda_{NC}$ } \label{ttp1}
 \end{figure}

\begin{figure}
 \includegraphics[scale = 0.56]{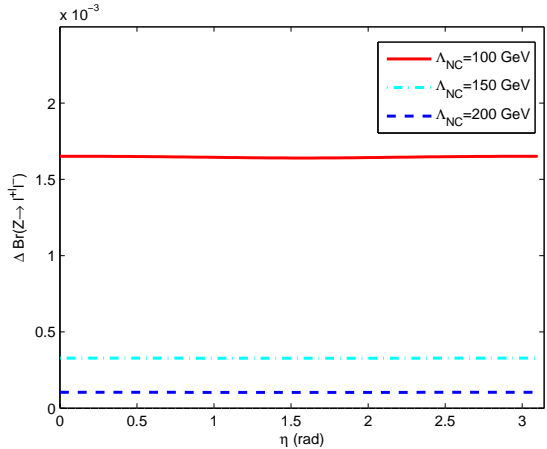}%
\caption {The $\eta$ dependence of $\Delta BR$($Z\rightarrow
e^{+}e^{-}$) for $\kappa=1$ } \label{ttp2}
 \end{figure}

Due to the Earth's rotation, any observable calculated in the NC
space-time frame should depend on time. On the other hand, it is
difficult to follow the experiments in time. It is therefore
reasonable to average the cross section or decay width over a full
day. For our problem, the time-averaged decay width is
\begin{equation}
\langle\Gamma\rangle_{T}=\frac{1}{T_{day}}\int_{0}^{T_{day}}dt\int_{-1}^{1}d(\cos\theta)\int_{0}^{2\pi}d\phi
\frac{d\Gamma}{d\cos\theta d\phi}.
\end{equation}
In particular, we are interested in the NC correction of the branch ratio
\begin{equation}
\Delta\textup{BR}
=\frac{\delta\Gamma}{\Gamma_{0}}\equiv\frac{\Gamma-\Gamma_{\textup{SM}}}{\Gamma_{\textup{SM}}}.
\end{equation}
The behavior of $\Delta \textup{BR}$ for different $\kappa$, as well
as the current experimental uncertainty, is shown in Fig.\ \ref{ttp1}
as a function of the NC scale $\Lambda_{NC}$. In the numerical
analysis, we use the input parameters of Ref.\ \cite{Chong}. The
location and  orientation of laboratory frame are set to be
$(\delta, a) = (\frac{\pi}{4}, \frac{\pi}{4})$, where the LEP experiment
measures the decay width of $Z\rightarrow e^{+}e^{-}$. We can see
from  Fig.\ \ref{ttp1} that $\Delta \textup{BR}$ is sensitive to both
$\kappa$ and $\Lambda_{NC}$. Clearly, the NC correction is
significantly enhanced as $\Lambda_{NC}$ or $\kappa$ decreases.
Compared with the experimental branch ratio Br$(Z\rightarrow
e^{+}e^{-})=3.363\pm 0.004\%$, for the choice of $\kappa=1$, a bound
on the noncommutative scale $\Lambda_{NC}\geq 150$ GeV is obtained by
imposing the constraint $\Delta \textup{BR}\leq 3\times 10^{-4}$. As
seen from Eq.\eqref{c2}, the NC correction of decay width also
depends on the orientation of $\theta_{E}$ i.e., the parameter
$\eta$. In Fig.\ \ref{ttp1}, we have set $\eta=\frac{\pi}{2}$. It is
thus necessary to investigate the sensitivity of $\Delta \textup{BR}$
on $\eta$. The NC correction of the time-averaged decay width is
presented as the function of parameter $\eta$ in Fig.\ \ref{ttp2}.
One can see from Fig.\ \ref{ttp2} that the NC effect produces a
positive deviation from the SM branch ratio for the
whole range of $\eta$. Despite the fact that a slightly peaked distribution appears, the curve is not sensitive to $\eta$. In this
sense, the bound obtained from Fig.\ \ref{ttp1} should be credible.

\begin{figure}
 \includegraphics[scale = 0.56]{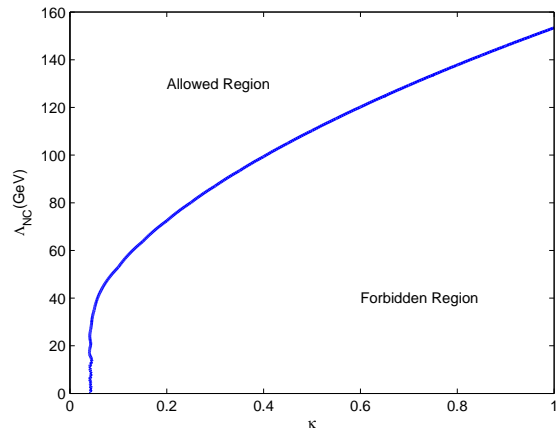}%
\caption {Bound on $\Lambda_{NC}$ as the function of $\kappa$ where
we set the range of $\kappa$ at [0,1]} \label{ttp3}
 \end{figure}

In Fig.\ \ref{ttp3}, we show the allowed region of $\Delta $BR  in
the $(\kappa, \Lambda_{NC})$ plane for $-1<\kappa<1$. We see that as
$\kappa$ increases, a higher bound on the NC scale appears.
Furthermore, a lower limit $\kappa\geq0.04$ is found for the
forbidden region when we set $\kappa$ to zero. This means that in
Eq.\ \eqref{rule}, for $\kappa=0$ the NC correction to the magnitude
of $Z\rightarrow e^{+}e^{-}$ only appears as a phase factor,
indicating that there is no NC deviation to the decay
width. Thus if we assume that the discrepancy between the
experimental and the SM results is fully induced by
the noncommutative effect, the value of $\kappa$ can not be
arbitrarily small. In the hybrid feature, additional sin-type
deformation shows up in the Feynman rule of NC $Z-e-e$ interaction
and leads to NC correction which is potentially detectable
or allows us to set bound on the NC parameters in high-accuracy
measurements of $Z$ decay width.

In conclusion, the $Z\rightarrow e^{+}e^{-}$ channel provides an
ideal process to understand not only the space-time noncommutativity,
but also the mathematical structure of the corresponding gauge theory. We
showed that the decay width is sensitive to both $\Lambda_{NC}$ and
the parameter $\kappa$ for the freedom of the hybrid gauge
transformation. %%Our study was carried out in the laboratory frame to get the credible results before we fit them to the experimental value. (Removed. This sentence has no meaning anyway.)
In terms of the NC effect, the discrepancy between the experimental
and SM results allows us to set a bound on the noncommutative
parameters. Although the current experimental uncertainty is still a little
large, the next generation $Z$ factory with the
Giga-$Z$ option of the International Linear Collider can generate
$2\times10^{9}$ $Z$ events at resonance energy\cite{Giga1,Giga2}. We
therefore expect that the high-luminosity $Z$ factory can
significantly enhance the sensitivity to probe the noncommutative
model via $Z$ boson decays.

\begin{acknowledgments}
The authors would like to thank Huang-sheng Xie and M. Y. Yu for useful
discussion during this work. This work is supported by the
Fundamental Research Funds for the Central Universities.
\end{acknowledgments}

\end{document}